# High pressure induced precipitation in Al7075 alloy

Abhinav Parakh[1], Andrew C. Lee[1], Stella Chariton[2], Melody M. Wang[1], Mehrdad T. Kiani[1], Vitali B. Prakapenka[2], and X. Wendy Gu[3]*

[1]Department of Materials Science and Engineering, Stanford University, Stanford CA

[2]Center for Advanced Radiations Sources, The University of Chicago, Chicago, Illinois 60637, USA

[3]Department of Mechanical Engineering, Stanford University, Stanford CA

*Corresponding author – xwgu@stanford.edu

**Abstract:** Precipitate-matrix interactions govern the mechanical behavior of precipitate strengthened Al-based alloys. These alloys find a wide range of applications ranging from aerospace to automobile and naval industries due to their low cost and high strength to weight ratio. Structures made from Al-based alloys undergo complex loading conditions such as high strain rate impact, which involves high pressures. Here we use diamond anvil cells to study the behavior of Al-based Al7075 alloy under quasi-hydrostatic and non-hydrostatic pressure up to ~53 GPa. In situ X-ray diffraction (XRD) and pre- and post-compression transmission electron microscopy (TEM) imaging are used to analyze microstructural changes and estimate high pressure strength. We find a bulk modulus of 75.2 ± 1.9 GPa using quasi-hydrostatic pressure XRD measurements. XRD showed that non-hydrostatic pressure leads to a significant increase in defect density and peak broadening with pressure cycling. XRD mapping under non-hydrostatic pressure revealed that the region with the highest local pressure had the greatest increase in defect nucleation, whereas the region with the largest local pressure gradient underwent texturing and had larger grains. TEM analysis showed that pressure cycling led to the nucleation and growth of many precipitates. The significant increase in defect and precipitate density leads to an increase in strength for Al7075 alloy at high pressures

## Introduction

Aluminum alloys are extensively used in aerospace, automobile, naval and construction applications [1] due to their high strength to weight ratio and corrosion resistance, and low cost [2–5]. Material properties such as strength, wear, and other tribological properties can be further improved through reinforcement with $Al_2O_3$, $B_4C$, $SiC$, $TiO_2$, and C particulates [6–12]. Precipitate strengthened Al7075 is a common high strength aerospace grade Al alloy, that can be heat-treated to achieve desired properties. Al7075 alloy is formed by alloying Al with Mg, Zn, and Cu. The T6 temper heat treatment results in the maximum ultimate tensile strength and yield strength [4]. T6 temper involves solution heat treatment at 470ºC for 1 hour followed by cold water quenching and then aging at 120ºC for 24 hours [4]. Different types of precipitates, including coherent metastable Guinier–Preston (GP) zones, metastable semicoherent η' phase precipitates, and incoherent η phase precipitates, are formed during heat treatments [13–15]. The precipitates contribute to strengthening of the Al7075 alloy, which is governed by the Orowan dislocation bypassing mechanism, in addition to solid solution strengthening due to Mg, Zn, and Cu atoms, and strengthening due to dislocation interactions [16]. The Al matrix has high stacking fault energy which makes the formation of stacking faults and twins unfavorable over the formation of full dislocations and dislocation loops, which changes the mechanical properties of Al-based alloys when compared to low stacking fault based alloys [17].

As with other structural alloys, the mechanical properties of Al7075 are closely tied to microstructure, which can change dramatically during processing [18,19]. Ma et al. found that ultra-fine grain Al7075 alloy has higher strength than coarse-grained Al7075 for equivalent heat treatments [20]. Shaeri et al. showed that Al7075 alloy undergoing severe deformation by equal channel angular pressing (ECAP) has improved mechanical properties, and ECAP processing accelerated the formation of precipitates in the alloy [19]. Kumar et al. showed that ECAP breaks down rod-like $MgZn_2$ precipitates into smaller spherical precipitates [21]. Researchers have observed deformation induced precipitation under high temperature deformation in Al7xxx alloys [22,23]. Precipitate formation and redissolution into the matrix is a dynamic process that governs the mechanical properties of Al7075 alloys and it is important to understand how stress state can affect the precipitates.

Laser shock peening, shot peening, and deep rolling have been used to improve the fretting fatigue, strength, and hardness of Al7075 alloy [24,25]. These structural and mechanical changes are particularly challenging to understand during shock, impact, and ballistic loading, yet this is critical for the use of Al7075 in armor, munitions, and military aircraft [26,27]. These loading conditions include complex stress states, pressures of tens of GPa to Mbar, and strain rates of $10^2$/s to $10^6$/s [28]. In situ studies, in which the evolution of microstructure and properties is monitored during loading, are particularly difficult due to the short time scale of these experiments. Here, we use high pressure diamond anvil cell (DAC) techniques to compress Al7075 under quasi-hydrostatic and non-hydrostatic stress states to tens of GPa pressures. DAC is a static experiment that allows for accurate in situ measurements using X-ray or optical scattering or spectroscopy. In situ X-ray diffraction (XRD) is used to reveal microstructural changes with pressure including strain field, defect formation, and grain growth under pressure [29]. The results of the DAC experiments can be compared to shock and high-strain rate testing to separate high pressure and strain rate effects, and guide the design of alloys for extreme conditions. Alternatively, it is possible that the high pressure DAC measurements could be used as a screening step for identifying materials for extreme conditions, which would then be subjected to more challenging high strain rate experiments.

We studied T6 tempered Al7075 alloy under quasi-hydrostatic pressure up to 53 GPa and under non-hydrostatic pressure up to ~38 GPa. The bulk modulus of Al7075 alloy was measured to be 75.2 ± 1.9 GPa using XRD peak position under quasi-hydrostatic conditions. Under non-hydrostatic pressure, the XRD peak width showed that the Al7075 alloy is undergoing defect formation with increasing pressure. XRD mapping was performed under non-hydrostatic pressure showing the variation in microstructure at different radial distances from the center of the sample chamber. XRD mapping showed that pressure cycling led to the strengthening of the alloy. Pre-compression and post-compression samples were observed using transmission electron microscope (TEM) to track the microstructural changes with pressure cycling. TEM images showed that pressure cycling made Mg and Zn atoms unstable in solid solution with the Al matrix. Post-compression GP zones and η' phase precipitates grew significantly in number density when compared to pre-compression. In situ XRD measurements were conducted which showed that the precipitates grow under high pressure. η' phase precipitate growth was due to formation and agglomeration of GP zones with increasing pressure. The formation of new η' phase precipitates and GP zones in addition to dislocation nucleation and interaction led to the strengthening of the alloy with pressure cycling.

**Experimental Methods**

*Materials*

400 μm thick T6-Al7075 alloy sheet was purchased from Goodfellow and was polished to a metallographic finish. Toluene was purchased from Sigma-Aldrich. 250 μm thick Re metal was purchased from Thermo-Fischer Scientific.

*High Pressure X-ray Diffraction*

High pressure experiments were performed using two symmetric Mao-type DACs with 300 μm and 400 μm culets. Rhenium metal was used as a gasket for high pressure experiments. Re gasket was indented to a thickness of 50 μm and then an electric discharge machining was used to drill the sample chambers. Neon was loaded using a gas-loading system at GSECARS (Argonne National Laboratory, APS, USA) and was used as a quasi-hydrostatic pressure medium in the 300 μm DAC up to ~53 GPa. Toluene was used as a non-hydrostatic pressure medium in the 400 μm DAC up to ~38 GPa [30]. A small piece of the T6-Al7075 alloy was loaded into the DACs for pressurization. Ruby powder was loaded along with the sample to determine the pressure using ruby fluorescence [31]. The extent of non-hydrostatic stress state in the DAC was inferred using the peak broadening and splitting of ruby fluorescence peak [32–34]. In situ XRD experiments were performed at the 13-IDD beamline at GSECARS (Advanced Photon Source, APS, USA). A membrane pressurization system was used to control the pressure of the cell remotely for the quasi-hydrostatic DAC, and the pressure was manually adjusted using screws for the non-hydrostatic DAC. XRD experiments were performed using a monochromatic X-ray beam with the energy 42 keV and a wavelength of 0.2952 Å. Dectris Pilatus CdTe 1M was used as the detector. Diffraction patterns were collected for 30 s or 60 s depending upon the pixel saturation on the detector. The sample to detector distance was calibrated using NIST LaB$_6$ standard. The 2D images were integrated to 1D plots using DIOPTAS software [35]. The XRD peak parameters were calculated by fitting the peaks to a combination of Gaussian and Lorentzian peak profiles along with a high order polynomial for the background using OriginPro software. Rietveld refinement was performed using X'Pert Highscore Plus software with Re and Al$_{0.9}$Mg$_{0.03}$Zn$_{0.07}$ as the phases with spherical harmonics to model texture within the material.

*Microstructure Analysis*

Ambient pressure XRD was collected at the same X-ray beam and detector conditions as the high-pressure XRD. Ambient pressure XRD was collected by placing a small piece of T6-Al7075 alloy on a 50 μm thick Kapton tape and the diffraction was collected for 15 s. Background diffraction signal was collected through just the Kapton tape for 15 s. The optical micrographs were collected using a Nikon upright microscope with a X20 and X100 objectives. DAC with 400 um culet was used to compress T6-Al7075 alloy in toluene non-hydrostatic medium to 30 GPa pressure to prepare a sample for TEM characterization. TEM lamellae were made using FEI Helios NanoLab 600i Dual Beam scanning electron microscope/focused ion beam system from pre-compressed and post-compressed Al7075 alloy. TEM was performed using FEI Tecnai G2 F20 X-TWIN at 200 keV accelerating voltage. TEM images were analyzed using ImageJ software.

**Results and Discussion**

*Pre-compression microstructure analysis*

Mg, Zn and Cu are the main alloying elements in Al matrix with a composition of 2.1-2.9 wt%, 5.1-6.1 wt%, and 1.2-2 wt%, respectively [36,37]. Fig. 1A shows the unrolled 2D XRD image of Al7075 alloy collected at ambient pressure. The diffraction pattern starts from 0º to 360º azimuthal angle on the detector and has several regions with no pixels shown in black parabolic curves and lines due to detector artifacts. These regions were masked while integration was performed to generate a 1-D diffraction pattern. Several straight diffraction lines are observed in Fig. 1A which correspond to the Al matrix diffraction.

Intensity along the azimuthal angle is uniform for each diffraction line indicating that the sample was not textured along the azimuthal angle. A weak non-uniform diffraction line at $2\theta$ of 7.7° and a sharp diffraction spot at $2\theta$ of 13° corresponds to the $MgZn_2$ precipitate. Fig. 1B shows the integrated 1-D XRD pattern for T6-Al7075 alloy. The FCC peak pattern for the Al matrix results in (111), (200), (220), (311), and (222) peaks, as well as higher order diffraction peaks. The lattice parameter for Al matrix measured from the diffraction pattern was 4.053 Å which is close to the value reported for Al7075 alloy and the value for bulk Al [15]. The average grain size calculated from Williamson-Hall analysis using XRD peak widths and positions was $0.7 \pm 0.05$ μm and the microstrain was $0.18 \pm 0.02\%$ due to material processing conditions. Low intensity $MgZn_2$ peaks were also observed which correspond to the η' phase precipitates.

Mg and Zn precipitate out from the Al matrix solid solution upon different aging treatments [4,20]. Mg and Zn segregate first into a coherent GP zone. Next, they grow into a semicoherent η' phase $MgZn_2$ precipitate, and then into an incoherent η phase $MgZn_2$ precipitate [19]. Fig. 2A and 2B show clusters of several GP zones in TEM. GP zones generally segregate and cluster together near a grain boundary [14,19]. We were not able to observe grain boundaries in TEM, possibly because the high concentration of GP zones obscured these features. The average size of GP zones was $5.6 \pm 1.6$ nm (81 GP zones analyzed). Fig. 2C shows a higher magnification image of GP zones showing that the spacing between the GP zones is about $20 \pm 7$ nm. Fig. 2D shows a high resolution TEM image of several GP zones with visible lattice planes, in which the GP zones have the same crystal orientation and have a coherent boundary with the Al matrix. Nucleation of a GP zone is preferred because precipitates are only a few nanometers in size and form a coherent boundary with the Al matrix to minimize the surface energy at the expense of volumetric lattice strain. Fig. 2E and 2F show the larger sized $MgZn_2$ η' phase precipitate which has a semicoherent boundary with the Al matrix. There were only a few η' phase precipitates observed, and the average size of the precipitates was $77 \pm 8$ nm (4 precipitates were analyzed).

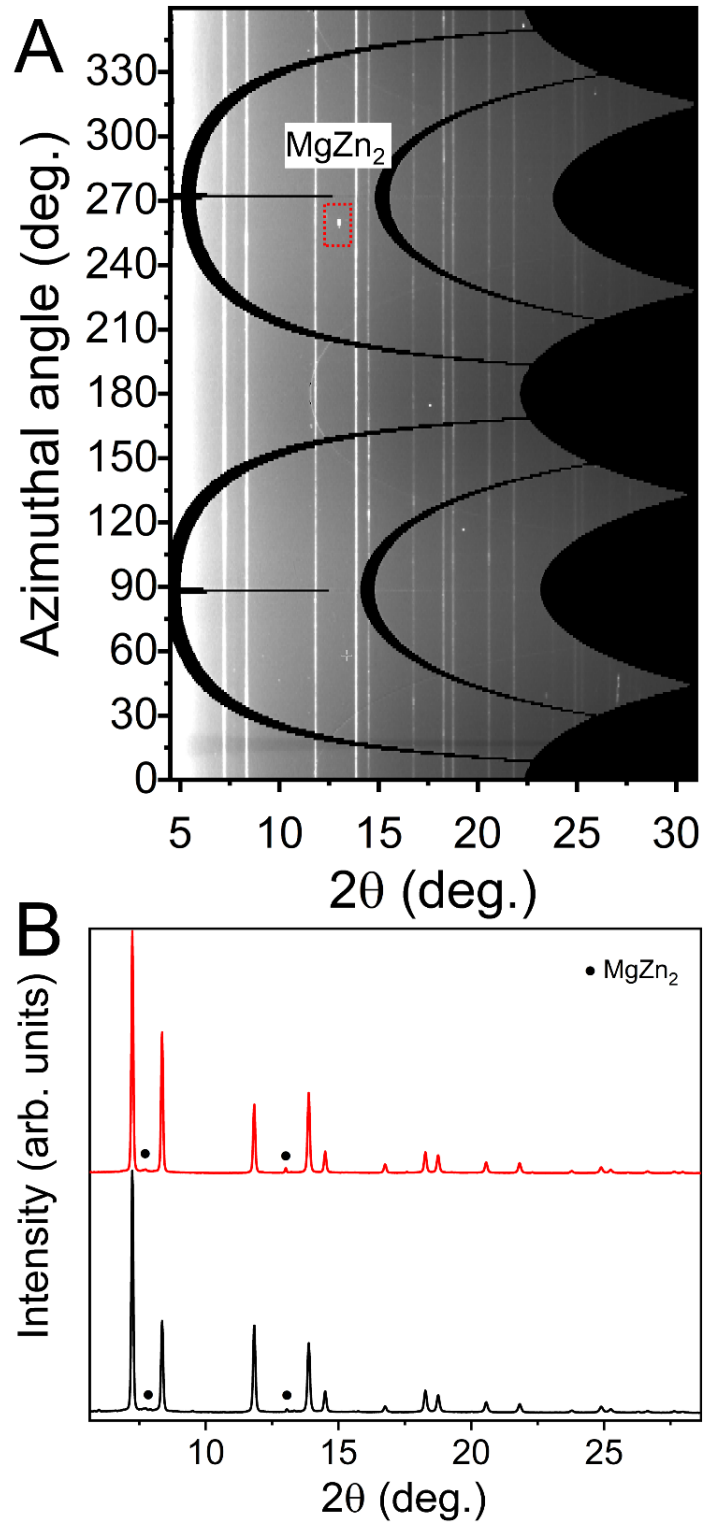

**Figure 1: A) Ambient pressure 2D XRD cake pattern showing the textured $MgZn_2$ precipitate pattern (in red box).** B) Ambient pressure 1D XRD for T6-Al7075 alloy at two different sample positions showing the sharp Al FCC matrix peaks (continuous lines) and $MgZn_2$ precipitate peaks.

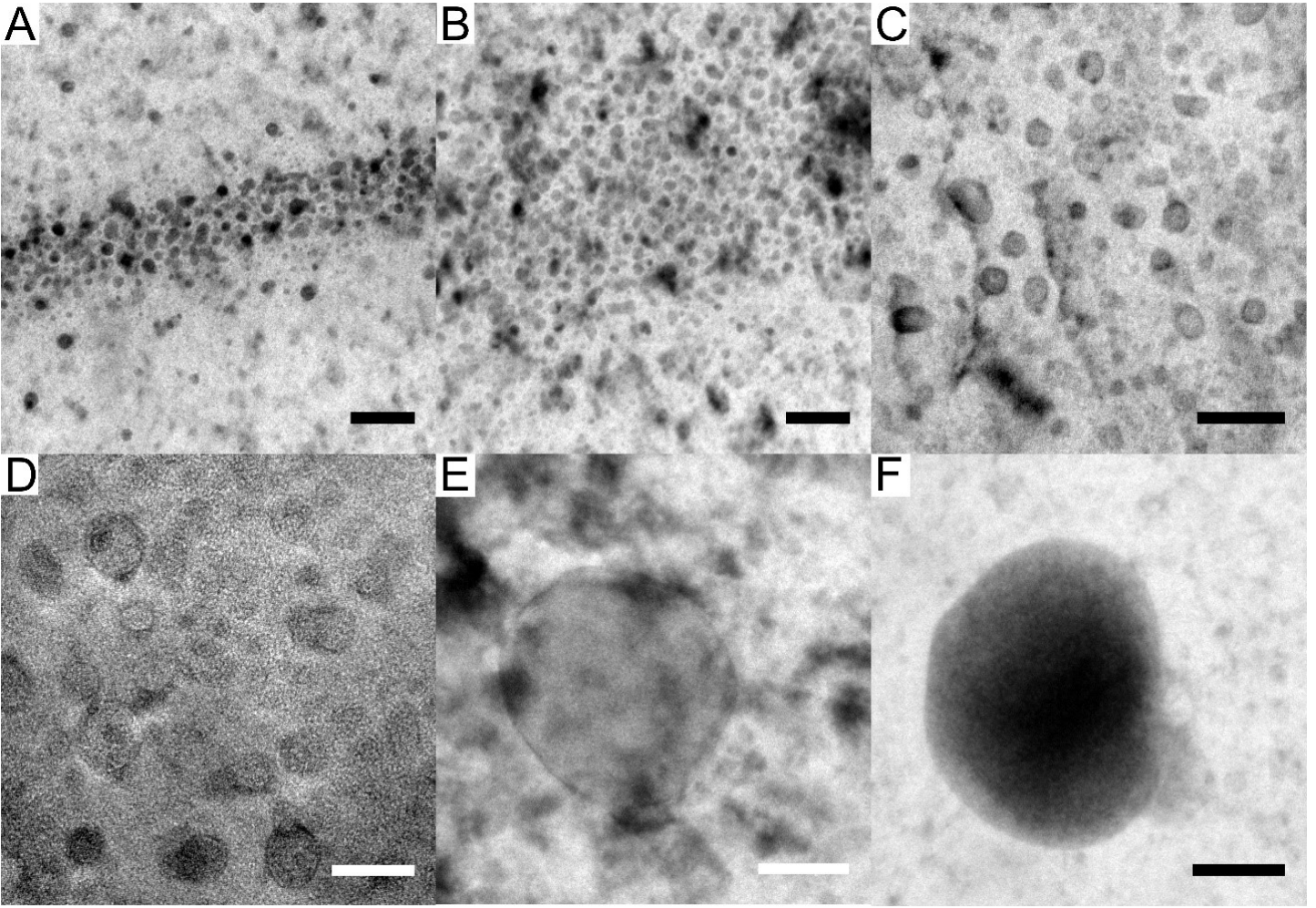

**Figure 2: TEM images of pre-compressed T6-Al7075 alloy.** A) Linear band of GP zones. B) Cluster of GP zones. C) and D) Higher magnification TEM images of GP zone clusters showing the coherent interface between GP zones and Al matrix. E) and F) Bright field TEM image of η' phase MgZn$_2$ precipitate. Scale bars are 25 nm for A-C, E and F, and 10 nm for D.

*High pressure deformation*

Quasi-hydrostatic in situ XRD measurements were performed using neon as a pressure medium. XRD patterns were collected while increasing sample chamber pressure at 7.4 GPa, 16.8 Gpa, 27 Gpa, 35.3 Gpa, 42.0 Gpa, and 53.0 Gpa, and after unloading the membrane system at 37 Gpa pressure. Fig. 3A shows the fitted diffraction pattern with increasing pressure under quasi-hydrostatic stress state. Al (111), (200), (220) and (311) diffraction peaks were visible with increasing pressure. The peaks shifted to higher 2θ angles with increasing pressure. The (111) peak shifted by 12% at the maximum pressure which corresponds to ~29% volumetric compression, and recovered to 9% of the original value after unloading to 37 Gpa. The XRD peak positions were used to calculate the unit cell volume as a function of pressure and then used to fit the Birch–Murnaghan equation of state [38]:

$$P(V) = \frac{3B_0}{2}\left[\left(\frac{V_0}{V}\right)^{\frac{7}{3}} - \left(\frac{V_0}{V}\right)^{\frac{5}{3}}\right]\left\{1 + \frac{3}{4}(B_0' - 4)\left[\left(\frac{V_0}{V}\right)^{\frac{2}{3}} - 1\right]\right\}(1)$$

Where, $P$ is the hydrostatic pressure measured, $B_0$ is the bulk modulus, $B_0'$ is the derivative of the bulk modulus with respect to pressure, $V_0$ is the unit cell volume at ambient pressure, and V is the unit cell volume at pressure P. Fig. 3B shows the unit cell volume as a function of pressure and the corresponding Birch-Murnaghan equation of state fit. The calculated $B_0$ was $75.2 \pm 1.9$ Gpa which is close to the reported bulk modulus for Al matrix and $B_0'$ was $4.3 \pm 0.2$ Gpa [39]. The unload to 37 Gpa also followed the same equation of state (see Fig. 4B). Peak width as a function of pressure was tracked using a combination of Gaussian and Lorentzian peak profiles to fit the diffraction pattern. Fig. 3C shows the peak width for the (111) and (200) peak with pressure and shows the corresponding linear line fit for the peak width with increasing pressure. Peak width for (111) peak increased with pressure upon loading to ~50% at the maximum pressure. The (111) peak width increase could be due to the shift of the peak towards higher $2\theta$ angles, as well as the formation of twins or dislocations under the slight non-hydrostatic stress state in neon pressure medium under high pressure [40,41]. Upon unloading to 37 GPa, the peak width for (111) peak recovered back to 42% of the original value and followed closely the linear fit for peak width increase with increasing pressure. The peak width for (200) peak also increased with increasing pressure. The neon pressure medium diffraction peaks overlapped with the Al (200) peak at several pressures. Neon diffraction spots were identified and removed from the analysis to the best of our abilities but could still be convoluted with the Al (200) peak.

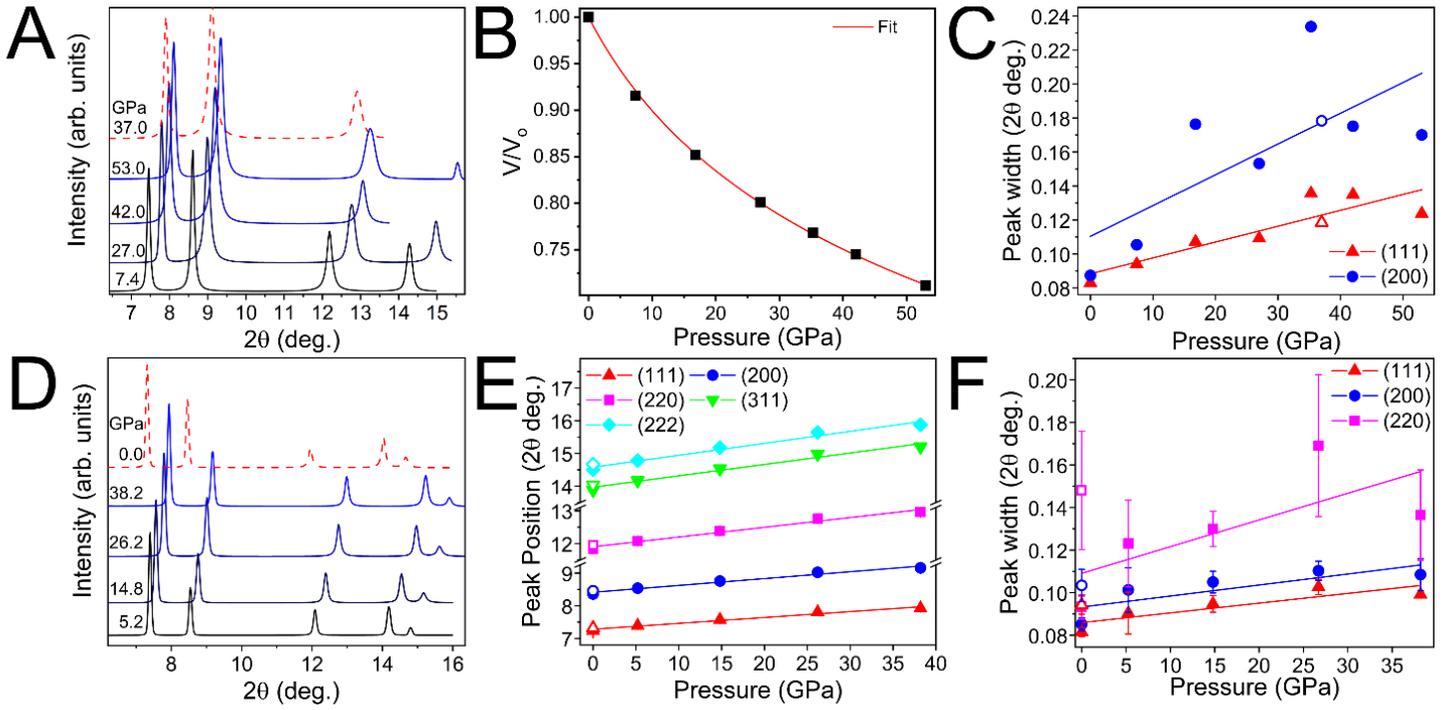

**Figure 3: High pressure XRD analysis.** A-C) Quasi-hydrostatic pressure environment and D-F) non-hydrostatic pressure environment. A) In situ XRD patterns with increasing pressure (solid lines). XRD with decreasing pressure is shown using red dashed line. B) Birch-Murnaghan equation of state fit. C) XRD peak width for (111) and (200) peaks with increasing pressure (solid symbol), and decreasing pressure (open symbol). D) In situ XRD patterns with increasing pressure (solid lines). XRD with decreasing pressure is shown using red dashed line. E) Overall peak position for XRD peaks with increasing pressure (solid symbol), and decreasing pressure (open symbol). F) Overall peak width change for (111), (200) and (220) peak with increasing pressure (solid symbol), and decreasing pressure (open symbol).

Non-hydrostatic pressure in situ XRD experiments were performed using toluene as the pressure medium. Toluene freezes around 1.9 GPa and then becomes a non-hydrostatic pressure medium which applies a larger force along the axis of the diamond than in the transverse direction [30]. The non-hydrostatic effect of the toluene pressure medium was observed on the ruby fluorescence peak width and peak splitting as the ruby R1 and R2 peak width increased and the peak distance varied from the ideal distance with increasing non-hydrostatic pressure [32–34]. Fig. 5A shows the fitted diffraction pattern with increasing pressure under non-hydrostatic stress state. XRD patterns were collected with increasing pressure at 5.2 GPa, 14.8 GPa, 26.7 GPa, and 38.2 GPa and after unloading at 0 GPa pressure. These reported sample chamber pressures are obtained from the ruby fluorescence and rhenium diffraction. Diffraction was collected at 7 different points on the non-hydrostatically compressed Al7075 sample. Al (111), (200), (220), (311) and (222) diffraction peaks were visible with increasing pressure (see Fig. 3D). The peaks shift to higher 2θ angles with increasing pressure; on average the (111) peak shifted by 9.4% at the maximum pressure which corresponds to ~24% volumetric compression, and the (111) peak position recovered to 1% of the original value after unloading to 0 GPa (see Fig. 3E). Fig. 3F shows the peak width with pressure for (111), (200) and (220) peak along with a linear line fit for the peak width with increasing pressure. (111) peak width increased by 21% at the maximum pressure and remained wider by 15% after unloading to 0 GPa pressure. The increase in the (111) peak width under non-hydrostatic stress was lower than under quasi-hydrostatic stress. This indicates that grain size remained larger under non-hydrostatic pressure, which would decrease the peak width, or that the neon diffraction peaks interfered with the sample peaks resulting in higher peak width for the hydrostatic case. Peak width reduction due to large grain size opposes the increase in peak width from increasing dislocation/defect density and the shift of the peak to a higher 2θ position [29]. After unloading, the peak width remained at a higher value due to the presence of higher dislocation/defect density. A permanent increase in the (200) and (220) peak width was also observed, which also indicates the formation of defects [42]. Precipitate peaks were difficult to observe in both quasi-hydrostatic and non-hydrostatic conditions due to their overlap with Re and Ne diffraction peaks.

Further insight into structural changes under non-hydrostatic pressure can be gained from XRD mapping at different locations on the sample. Seven diffraction spots were measured on the non-hydrostatically compressed sample. The average distance between the different points was $32 \pm 16$ μm and the minimum separation was ~13 μm. The X-ray beam spot size was smaller than 5 μm so there was no overlap between the different points. Fig. 4A shows the peak position of (111), (200) and (220) peak with radial distance from the center of the sample chamber at 38.2 GPa sample chamber pressure. (111), (200), and (220) peaks are shifted to higher 2θ values by an additional ~0.3% close to the center of the sample chamber compared to at the edge of the sample chamber. Fig. 4B shows the variation in peak position of the (111) peak with radial distance for 5.2 GPa, 14.8 GPa, 26.2 GPa, and 38.2 GPa sample chamber pressures. The variations between (111) peak position with radial distance were small for low sample chamber pressures (5.2 GPa and 14.8 GPa). For higher pressures, the peak position was generally shifted to higher 2θ angles for points closer to the sample chamber center. The difference between (111) peak position closer to the center and the furthest from the center increased by 36% with increasing sample chamber pressure from 26.2 GPa to 38.2 GPa.

The peak position shift at each point was used to calculate the local pressure at each point and is plotted with radial distance in Fig. 4C. We find that the local pressure is higher closer to the center of the sample chamber at 26.2 GPa and 38.2 GPa sample chamber pressure, and this difference increased with increasing sample chamber pressure from 26.2 GPa to 38.2 GPa. Fig. 4C also shows the linear line fits for pressure with radial distance from the center. The slope is zero for the 5.2 GPa and 14.8 GPa sample chamber pressure data. The slope of the linear line was used to roughly estimate the quasi-static yield strength of the alloy as a function of sample chamber pressure using the following equation [43,44] –

$$Y = 2\sigma_{rz} \approx -h\left(\frac{dP}{dr}\right) \qquad (2)$$

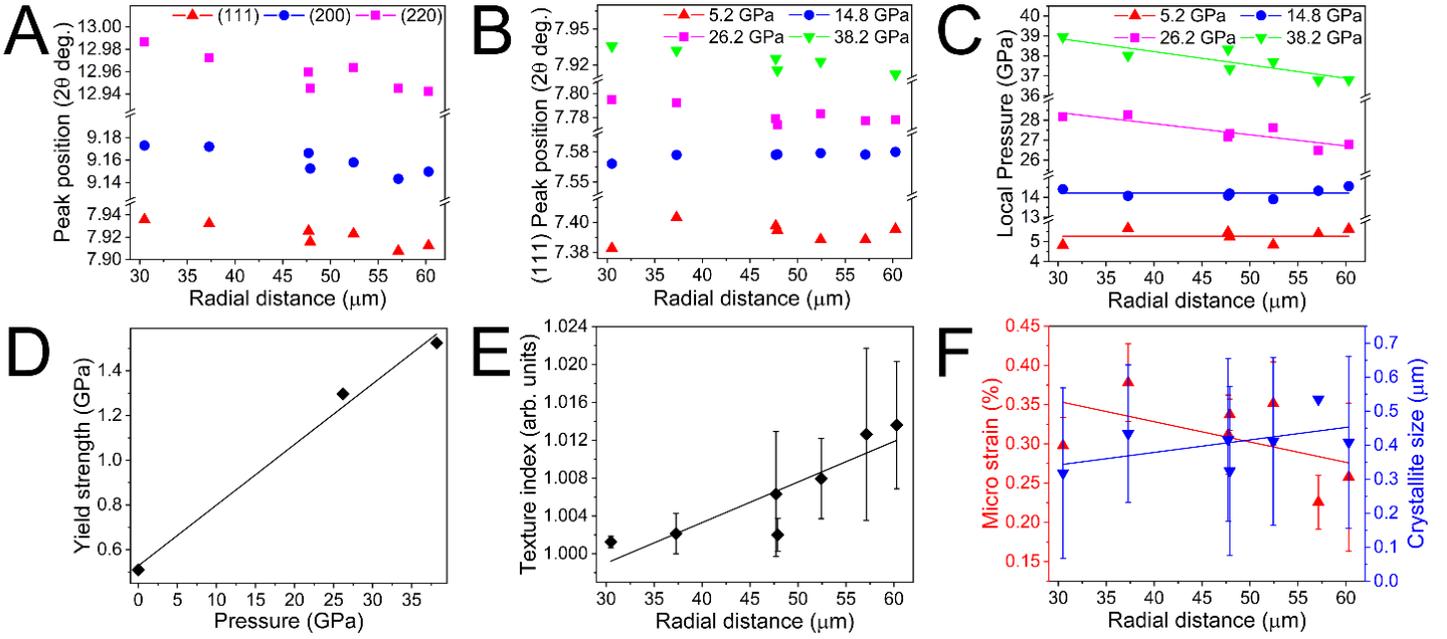

**Figure 4: Non-hydrostatic pressure XRD mapping.** A) Peak position for (111), (200) and (220) peak at different radial distance from the center of the sample chamber at 38.2 GPa. B) Peak position for (111) peak with increasing sample chamber pressure at 5.2 GPa, 14.8 GPa, 26.7 GPa and 38.2 GPa. C) Local pressure measured at each point using the change in lattice constant with radial distance at different sample chamber pressure (measured using ruby). D) Quasi-static yield strength with increasing pressure. The ambient pressure yield strength is the reported value from the manufacturer. E) Average texture index across all sample chamber pressures calculated using spherical harmonics preferred orientation model with radial distance. F) Average microstrain and crystallite size across all sample chamber pressures with radial distance.

Where, $Y$ is the quasi-static yield strength, $\sigma_{rz}$ is the shear stress, $h$ is the overall sample thickness, $P$ is the local pressure at each point, $r$ is the radial distance from the center of the sample chamber. The slope $\frac{dP}{dr}$ was estimated using the linear fit of pressure and radial distance (Fig. 6C). Sample thickness as a function of sample chamber pressure was estimated using the initial thickness (50 µm) and the bulk modulus of Al7075. Using this, we calculated the quasi-static yield stress at different sample chamber pressures. At a sample chamber pressure of 26.2 GPa, the yield stress was 1.3 GPa, and at a sample chamber pressure of 38.2 GPa, the yield stress was 1.5 GPa (see Fig. 4D). At 5.2 GPa and 14.8 GPa, the slope $\frac{dP}{dr}$ was zero which indicates that the sample had not plastically yielded. Ambient pressure yield strength was reported by the manufacturer to be 0.5 GPa. The yield strength of

Al7075 alloy increased with increasing sample chamber pressure.

Fig. 4E shows the texture index averaged across all sample chamber pressures. The texture index was calculated using spherical harmonics model for preferred orientation. The texture index closer to the center of the sample chamber remained very close to 1 (no texture) at all sample chamber pressures. The texture index was higher closer to the edge of the sample chamber. This indicates that the sample became textured at larger radial distances away from the center of the sample chamber. Fig. 4F shows the microstrain and crystallite size averaged across all sample chamber pressures. The microstrain and crystallite size were calculated from the XRD peak profiles using Rietveld refinement. The average microstrain decreases with increasing radial distance. The average crystallite size increases with increasing radial distance. Higher microstrain is related to higher defect density in the material. The region closer to the center of the chamber had more defects nucleated and smaller crystallite size than the region closer to the edge. A larger defect density closer to the center of the chamber was related to the higher local pressure closer to the center of the chamber whereas, the larger crystallite size closer to the edge of the chamber was related to a larger gradient in local pressure closer to the edge of the chamber. This matches the trend in texturing index as the grains closer to the center of the chamber remained randomly oriented with a high defect density and small crystallite size. The grains closer to the edge of the chamber were textured with a larger crystallite size and lower microstrain.

*Post-compression microstructure analysis*

T6-Al7075 sample was non-hydrostatically compressed to ~30 GPa and then analyzed using TEM (see Fig 5). Fig. 5A shows the low magnification TEM images of the compressed area. Fig. 5A shows that several large-sized η' phase precipitates of roughly 80

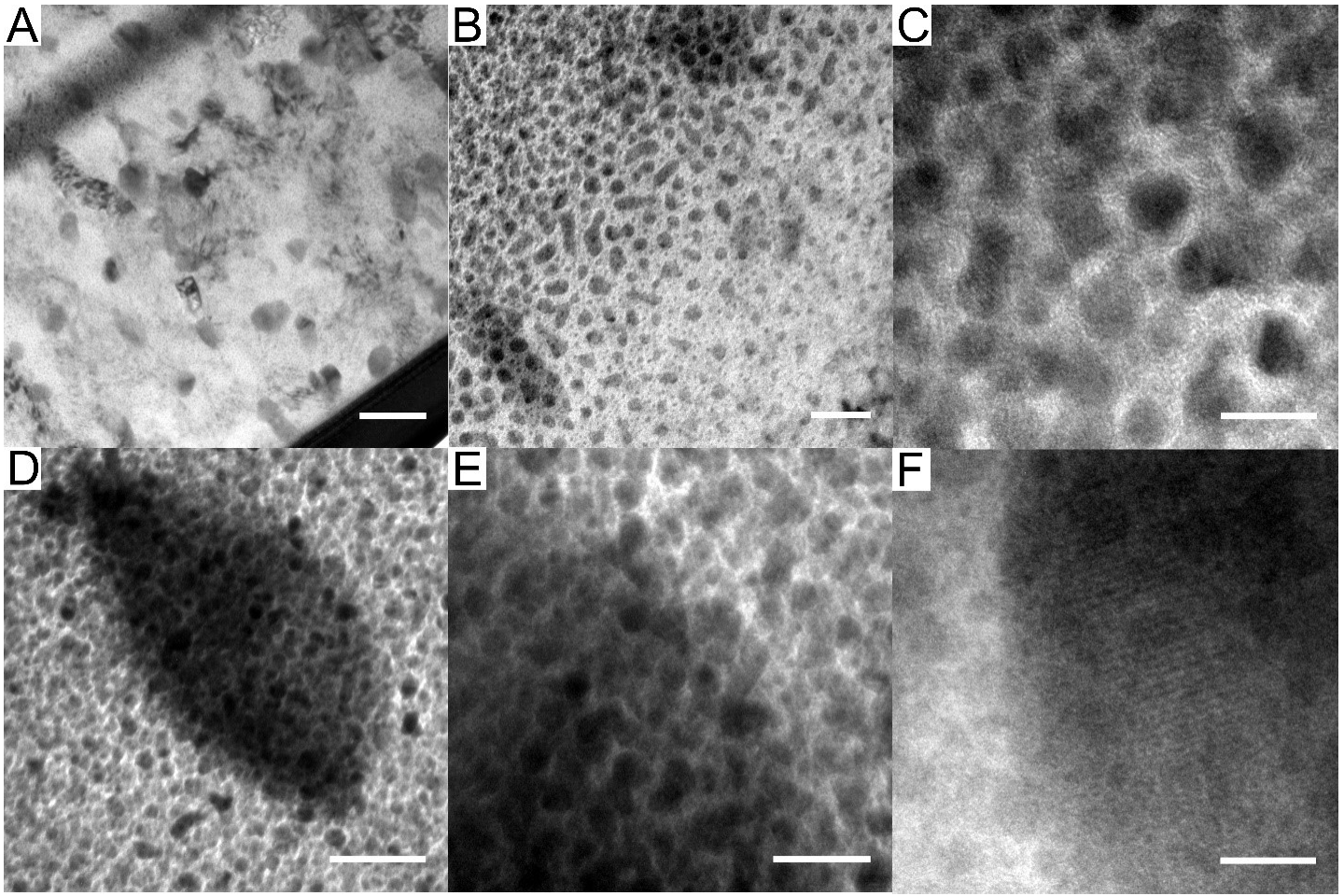

**Figure 5: TEM image of post-compressed T6-Al7075 alloy under non-hydrostatic condition.** A) Low magnification image showing the high density of large-sized η' phase precipitates. B) Higher magnification images showing the nucleation and aggregation of GP zones and smaller η' phase precipitates due to pressure cycling. C) Several GP zones and smaller η' phase precipitates with lattice planes in different directions. D) High density of GP zones and smaller η' phase precipitates near an incomplete large-sized η' phase precipitate. E) Pressure driven aggregation of GP zones and smaller η' phase precipitate to form a larger η' phase precipitate. F) High resolution image of a fully formed large-sized η' phase precipitate. Scale bars are 150 nm for A), 25 nm for B), 10 nm for C), 40 nm for D), 20 nm for E), and 10 nm for F).

± 20 nm in size (31 precipitates were analyzed) were formed due to pressure cycling. The density of η' phase precipitates after pressure cycling was ~16 precipitates/$\mu m^2$ which was considerably higher than pre-compression precipitate density (only a few precipitates were found across several $\mu m^2$ area). The average size of precipitates did not change significantly with pressure cycling. However, a few larger η' phase precipitates were observed in the compressed sample, which may indicate that precipitates grew in size compared to the pre-compressed sample. Fig. 5B-C shows high magnification TEM images of smaller-sized precipitates dispersed across the Al matrix. The lattice fringes for the smaller-sized precipitates were oriented in different directions indicating that these could be smaller η' phase precipitates or GP zones [20]. The GP zones and smaller η' phase precipitates were 6.0 ± 1.6 nm in size (500 precipitates were analyzed). A few elongated precipitates were observed which indicates growth through coalescence. Pre-compression GP zones were clustered around possible grain boundaries. Post-compression GP zones and smaller η' phase precipitates were evenly distributed and had a high density across the whole Al matrix. Precipitate density increase has previously been observed due to equal channel angular pressing processing [45].

Fig. 5D shows a TEM image of an incomplete large-sized η' phase precipitate with a high number density of GP zones and smaller η' phase precipitates around it. Higher magnification image at the boundary of the incompletely formed large-sized η' phase precipitate (Fig. 5D) is shown in Fig. 5E. GP zones and smaller η' phase precipitates migrate and aggregate to form an incomplete large-sized η' phase precipitate (Fig. 5E). The spacing between GP zones or smaller η' phase precipitates were still visible and individual small precipitates were separate from each other inside the forming large-sized η' phase precipitate. The overall density of GP zones or smaller η' phase precipitates increased significantly with pressure cycling. Fig. 5F shows the high resolution TEM image of a fully formed large-sized η' phase precipitate. Non-hydrostatic XRD results indicated that defects such as dislocations were nucleated due to pressure cycling. However, dislocations were not observed in the TEM samples because the large number of GP zones and smaller η' phase precipitates obscured the matrix.

Under high pressure, the Mg-Zn-Al solid solution that forms the Al7075 matrix becomes unstable due to the increased volumetric strain field associated with solid solution atoms. The Mg and Zn atoms precipitate out from the Al matrix by nucleating GP zones and smaller η' phase precipitates. High pressure also makes the metastable GP zones less favorable due to the increased volumetric strain of maintaining a coherent interface with the matrix so, smaller η' phase precipitates are most stable. We have observed a similar effect in 6 nm Au nanocrystals which transformed from multiply-twinned icosahedral to single crystalline particles due to the energetic cost of volumetric strain of multiply-twinned icosahedral shape under high pressure [46]. The non-hydrostatic pressure then drives the GP zones to aggregate and cluster together to form larger η' phase precipitates which releases the strain energy associated with solid solution atoms and the GP zones. Grain-boundary strengthening (Hall-Petch effect), dislocation strengthening (Bailey-Hrisch relationship), and precipitate strengthening (Orowan dislocation bypassing relationship) are the major strengthening mechanisms for Al7075 alloy at ambient conditions [16,47,48]. There is a minor contribution from solid-solution strengthening (Fleischer equation) [20]. At elevated non-hydrostatic pressures, increased dislocation/defect density (dislocation strengthening) and the growth of precipitates and high density of GP zones and smaller η' phase precipitates (precipitate hardening) contribute to the increase in strength [16]. Precipitation of Mg and Zn from the solid solution into GP zones and η' phase precipitates leads to increased dislocation interactions and pinning. This is observed as an increase in quasi-static yield strength with increasing pressure using in situ XRD.

**Conclusion**

In this work, we pressurized precipitate strengthened Al-based Al7075 alloy under quasi-hydrostatic and non-hydrostatic pressures of ~53 GPa and ~38 GPa, respectively using a diamond anvil cell. We used in situ XRD, and pre- and post-compression TEM imaging to understand structural changes and mechanical behavior of this alloy under such extreme conditions. The XRD patterns were analyzed by peak fitting and Rietveld refinement to understand the microstructural changes with pressure. XRD mapping over different regions on the sample was performed under non-hydrostatic conditions to understand the effects of local pressure and pressure gradient on the microstructure of the material. The following are the main conclusions from the study –

- Pre-compression TEM images showed that the Al7075 alloy had GP zones and large-sized semicoherent η' phase precipitates. The GP zones were clustered together in a band and only a few semicoherent precipitates were found.

- Bulk modulus was calculated using the Birch-Murnaghan equation of state under quasi-hydrostatic pressure. The bulk modulus for the alloy was $75.2 \pm 1.9$ GPa and the pressure driven derivative of the bulk modulus was $4.3 \pm 0.2$ GPa.

- Non-hydrostatic pressure cycling showed that the average XRD peak widths at different regions of the sample increased with increasing pressure and remained at higher values after unloading. This indicated that defects were nucleated in the sample with pressure cycling and led to peak broadening even with an associated increase in grain size.

- Non-hydrostatic pressure cycling showed that the local pressure is higher, and the gradient of local pressure is smaller closer to the center of the sample chamber than at the edge of the sample chamber. The region near the center of the sample chamber remained randomly oriented with pressure cycling with smaller grain size and higher microstrain due to larger defect density. The region near the edge of the sample chamber was textured and had a larger grain size and lower microstrain. The higher local pressure near the center of the sample chamber led to more defects being nucleated whereas, the larger gradient of local pressure near the edge of the sample chamber led to texturing with pressurization.

- Post-compression TEM images showed that a large number of precipitates were nucleated with pressure cycling. A high density of GP zones and semicoherent η' phase precipitates were observed after non-hydrostatic pressure cycling. The increase in the number of precipitates and higher number of defects led to the strengthening of the alloy at high pressures.

These results highlight the dynamic microstructure of precipitate strengthened Al-based alloys. This will enable engineers to design an optimum aging treatment for parts undergoing extreme conditions to achieve ideal strength and stability.

**Acknowledgement**

We thank Prof. Wendy Mao and Feng Ke at Stanford University for sharing their resources for drilling the sample chamber. We acknowledge the assistance by Dr. Radhika P. Patil in polishing the Al7075 samples. We also acknowledge the assistance of Sergey N. Tkachev with neon gas loading for quasi-hydrostatic diamond anvil cell measurements. X.W.G. and A.P. acknowledge financial support from the Army Research Office under grant number W911NF2020171. Part of this work was performed at GeoSoilEnviroCARS (The University of Chicago, Sector 13), Advanced Photon Source (APS), Argonne National Laboratory. GeoSoilEnviroCARS is supported by the National Science Foundation – Earth Sciences (EAR – 1634415) and Department of Energy-GeoSciences (DE-FG02-94ER14466). This research used resources of the Advanced Photon Source, a U.S. Department of Energy (DOE) Office of Science User Facility operated for the DOE Office of Science by Argonne National Laboratory under Contract No. DE-AC02-06CH11357. Use of the COMPRES-GSECARS gas loading system was supported by COMPRES under NSF Cooperative Agreement EAR -1606856 and by GSECARS through NSF grant EAR-1634415 and DOE grant DE-FG02-94ER14466. Part of this work was performed at the Stanford Nano Shared Facilities (SNSF), supported by the National Science Foundation under award ECCS-1542152. A.C.L is supported by Knight-Hennessy Scholars. M.M.W is supported by the NSF Graduate Fellowship. M.T.K. is supported by the National Defense and Science Engineering Graduate Fellowship. M.M.W and M.T.K were also supported by the grant DE-SC0021075 funded by the U.S. Department of Energy, Office of Science for TEM preparation and imaging.